%% file: main-arxiv.tex
\documentclass[sigconf,nonacm]{acmart}
\AtBeginDocument{%
  }

\usepackage{xspace}
\usepackage{enumitem}
\setlist[itemize]{noitemsep, topsep=0pt, partopsep=0pt, left=2pt, label=$\bullet$}
\usepackage{graphicx}
\usepackage{pifont}
\usepackage{makecell}
\usepackage{multirow}
\usepackage{amsmath}
\usepackage{booktabs}
\usepackage{soul,color,xcolor}
\usepackage{threeparttable} 

\usepackage[most]{tcolorbox}
\usepackage{refcount}  

\usepackage[T1]{fontenc}
\usepackage{marvosym}

\newcounter{finding}

\newtcolorbox{findingbox}[1][]{%
    colback=gray!10,
    colframe=black,
    boxrule=0.5pt,
    arc=2pt,
    boxsep=4pt,
    left=6pt,
    right=6pt,
    top=4pt,
    bottom=4pt,
    breakable,
    before skip=6pt,
    after skip=12pt,
    sharp corners=south,
    label={#1}
}

\newcommand{\finding}[2][]{%
    \refstepcounter{finding}%
    \begin{findingbox}[#1]%
         \noindent \textbf{\ensuremath{\blacktriangleright}~Finding~\Roman{finding}}: \textit{#2}
    \end{findingbox}%
}

\newtcolorbox{promptbox}[1]{
  colback=gray!5!white,
  colframe=gray!75!black,
  fonttitle=\bfseries,
  title=#1,
  sharp corners=southwest,
  boxrule=1pt,
  arc=4pt,
  left=6pt,
  right=6pt,
  top=6pt,
  bottom=6pt
}

\newcommand{\solidnum}[1]{%
  \ifcase#1\relax\or
  \ding{182}\or
  \ding{183}\or
  \ding{184}\or
  \ding{185}\or
  \ding{186}\or
  \ding{187}\or
  \ding{188}\or
  \ding{189}\or
  \ding{190}\or
  \ding{191}\else
  \textbf{#1}\fi
}

\newcommand{\LLMS}{LLMSE\xspace}
\newcommand{\LLMSs}{LLMSEs\xspace}
\newcommand{\eg}{\textit{e.g.}\xspace}
\newcommand{\ie}{\textit{i.e.}\xspace}

\newcommand{\uquery}{Rewitten queries\xspace}
\newcommand{\rlink}{{Retrieval references}\xspace}

\newcommand{\slink}{{Summary references}\xspace}

\newcommand{\rhit}{\textit{R-hit}\xspace}
\newcommand{\shit}{\textit{S-hit}\xspace}

\newcommand{\mseo}{LLMSEO\xspace}

\newcommand{\cp}[1] {{#1}}
\newcommand{\ww}[1] {{#1}}
\newcommand{\ignore}[1]{}
\begin{document}

\title{Unveiling the Resilience of LLM-Enhanced Search Engines against Black-Hat SEO Manipulation}

\author{Pei Chen}
\email{peichen19@fudan.edu.cn}
\affiliation{%
  \institution{Fudan University}
  \city{Shanghai}
  \country{China}
}

\author{Geng Hong \Letter}
\email{ghong@fudan.edu.cn}
\affiliation{%
  \institution{Fudan University}
  \city{Shanghai}
  \country{China}
}

\author{Xinyi Wu}
\email{xinyiwu20@fudan.edu.cn}
\affiliation{%
  \institution{Fudan University}
  \city{Shanghai}
  \country{China}
}

\author{Mengying Wu}
\email{wumy21@m.fudan.edu.cn}
\affiliation{%
  \institution{Fudan University}
  \city{Shanghai}
  \country{China}
}

\author{Zixuan Zhu}
\email{zhuzx24@m.fudan.edu.cn}
\affiliation{%
  \institution{Fudan University}
  \city{Shanghai}
  \country{China}
}

\author{Mingxuan Liu}
\email{liumx@mail.zgclab.edu.cn}
\affiliation{%
  \institution{Zhongguancun Laboratory}
  \city{Beijing}
  \country{China}
}

\author{Baojun Liu}
\email{lbj@tsinghua.edu.cn}
\affiliation{%
  \institution{Tsinghua University}
  \city{Beijing}
  \country{China}
}

\author{Mi Zhang}
\email{mi_zhang@fudan.edu.cn}
\affiliation{%
  \institution{Fudan University}
  \city{Shanghai}
  \country{China}
}

\author{Min Yang \Letter}
\email{m_yang@fudan.edu.cn}
\affiliation{%
  \institution{Fudan University}
  \city{Shanghai}
  \country{China}
}

\thanks{\Letter \ Corresponding authors.\\
To appear in Proceedings of the ACM Web Conference 2026 (WWW 2026).}

\renewcommand{\shortauthors}{Pei Chen et al.}

\begin{abstract}
\input{section/0-Abstract}
\end{abstract}



\maketitle

\input{section/1-Introduction}

\input{section/2-Background}
\input{section/3-Ecosystem}
\input{section/4-Traditional-Eva}

\input{section/5-New-Attack}
\input{section/6-Disscussion}
\input{section/8-Conclusion}
\begin{acks}
\input{section/Acknowledge}
\end{acks}

\bibliographystyle{ACM-Reference-Format}
\bibliography{Reference}

\appendix
\input{section/Appendix}

\end{document}

%% file: section/0-Abstract.tex
The emergence of Large Language Model-enhanced Search Engines (\LLMSs) has revolutionized information retrieval by integrating web-scale search capabilities with AI-powered summarization. While these systems demonstrate improved efficiency over traditional search engines, their security implications against well-established black-hat Search Engine Optimization (SEO) attacks remain unexplored. 

In this paper, we present the first systematic study of SEO attacks targeting \LLMSs. Specifically, we examine ten representative \LLMS products (e.g., ChatGPT, Gemini) and construct SEO-Bench, a benchmark comprising 1,000 real-world black-hat SEO websites, to evaluate both open- and closed-source \LLMSs. Our measurements show that \LLMSs mitigate over 99.78\% of traditional SEO attacks, with the phase of retrieval serving as the primary filter, intercepting the vast majority of malicious queries.
We further propose and evaluate seven \mseo attack strategies, demonstrating that off-the-shelf \LLMSs are vulnerable to \mseo attacks, i.e., rewritten-query stuffing and segmented texts double the manipulation rate compared to the baseline.
This work offers the first in-depth security analysis of the \LLMS ecosystem, providing practical insights for building more resilient AI-driven search systems.
We have responsibly reported the identified issues to major vendors.

%% file: section/1-Introduction.tex
\section{Introduction}

Among the emerging applications of large language models~(LLMs), the large language model-enhanced search engine (\LLMS) combines the vast search capabilities of the Web with efficient and precise responses to user queries. Due to their objectivity and ability to synthesize information, \LLMSs are increasingly being regarded as alternatives to traditional search engines.
For instance, Perplexity,
raise funds at an \$18 billion valuation in early 2025~\cite{perplexity2025valuation}. 

\begin{figure}[t]
    \centering
    \includegraphics[width=\linewidth]{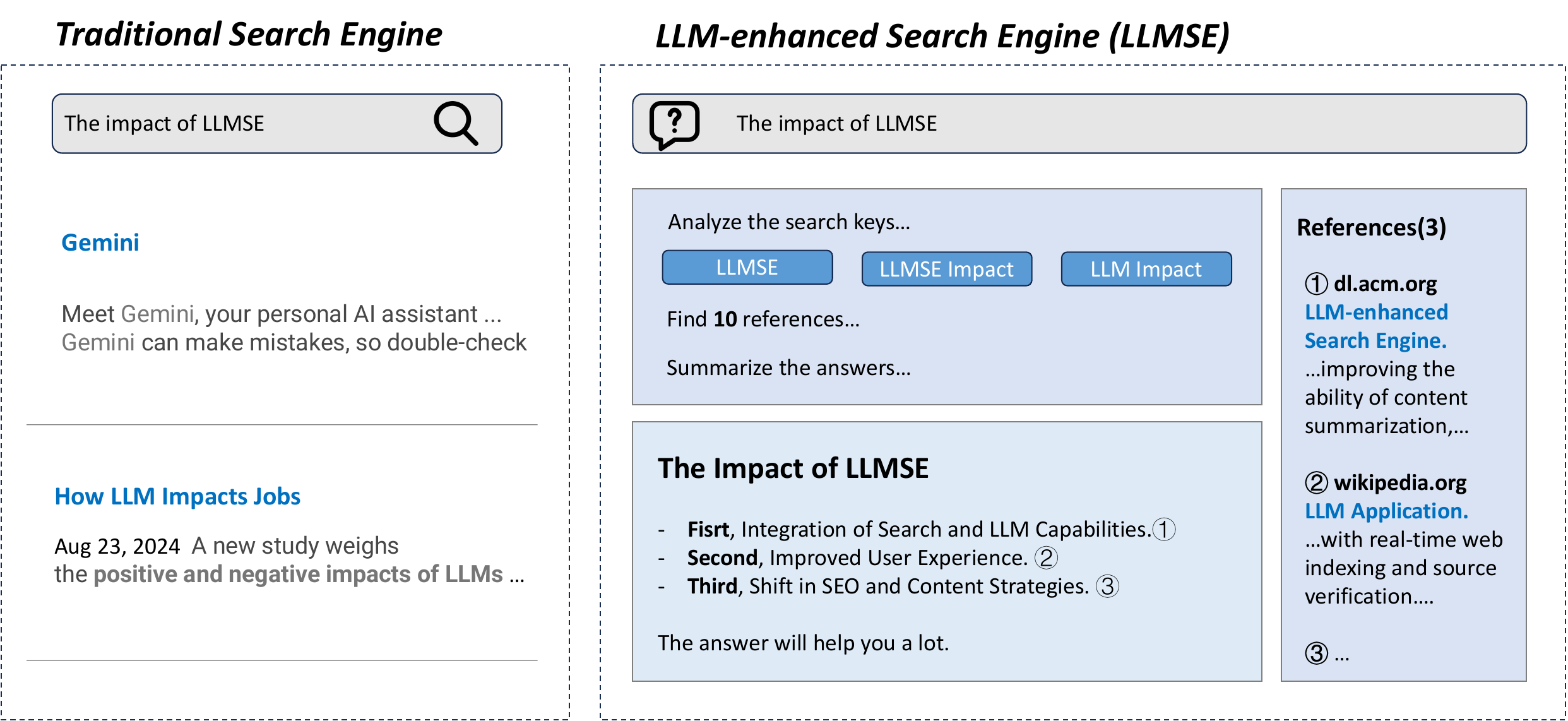}
    \caption{Examples of Traditional Search Engines vs. LLM-Enhanced Search Engines (\LLMS).}
    \Description{A comparison figure between traditional search engines and LLM-Enhanced search engines.}
    \label{fig:llms-example}
\end{figure}

Figure~\ref{fig:llms-example} illustrates a comparison between \LLMS and traditional search engines. The user begins by inputting a query to a practical problem, such as ``Impact of \LLMS?''. The traditional search engines return several separate Web sources of information, \eg news, forums.
while \LLMS directly generates a well-structured response 
providing clearer information in a well-defined overview. 

\cp{Despite these advantages, are \LLMSs truly more reliable than traditional search engines?}
\cp{
Search engine optimization~(SEO)~\cite{seo_wikipedia}, including black-hat SEO~\cite{Wu2005linkfarm, Ntoulas2006stuffing, Nektarios2011redirection, McCreadie2012contentfarm, liao2016characterizing, yang2021scalable}, has damaged search result quality on the traditional search engines for decades. 
With many attackers now turning to \LLMSs, likely reusing established SEO methods and even inventing new manipulations, the emerging \LLMS systems are facing a qualitatively new threat, underscoring the urgency of understanding and mitigating such risks.
}






\noindent\textbf{Research Gap.} 
%
Despite the growing deployment of \LLMSs, their security under SEO manipulation remains insufficiently understood.
On one hand, as a rapidly emerging field, most work on \LLMSs focuses on improving search efficiency, accuracy, and verifiability~\cite{spatharioti2023comparing, liu2023evaluating, wu2024easily, luo2025unsafe}. On the other hand, the security-related works pay attention to the model-level attack techniques\cite{nestaas2024adversarial,aggarwal2024geo,pfrommer2024ranking,zou2024poisonedrag}. 
For example, PoisonedRAG~\cite{zou2024poisonedrag} manipulates RAG results by poisoning knowledge databases, and GEO~\cite{aggarwal2024geo} tries to craft text-optimization attacks to gain visibility in the model summary.
However, these methods are typically evaluated in carefully crafted experimental settings, only partial components of the \LLMS workflow, neglecting their impact on the end-to-end system in real-world scenarios.
\cp{
\citet{luo2025unsafe} observed the harmful content and malicious URLs from \LLMS.
However, they do not analyze these threats from the perspective of traditional SEO techniques or assess their phase-specific vulnerabilities.
}
%
%
Moreover, \ww{how traditional black-hat SEO threats affect \LLMS} remain underexplored, posing persistent and transferable risks to evolving \LLMS infrastructures.

\noindent\textbf{Our Work.}
We conduct the first \ww{systematic} study to investigate the black-hat SEO threat to \LLMS. We try to answer the following three important research questions~(RQ):

\textit{\textbf{RQ1}: What is \LLMS workflow and whether the design of \LLMS is inherently resistant to SEO manipulations?}

\textit{\textbf{RQ2}: Will black-hat SEO attacks on traditional search engines affect \LLMS? If so, how do they affect each phase?}

\textit{\textbf{RQ3}: Are there any \mseo techniques that can significantly manipulating \LLMS results?}



Driven by these RQs, we first investigated 10 popular \LLMS products and analyzed the special workflow of \LLMS, revealing the attack surfaces. 
Second, we examined the effectiveness of traditional black-hat SEO techniques on \LLMSs. 
We constructed SEO-Bench with 1,000 real-world black-hat SEO attacks and then evaluated the defense performance of open-source and closed-source \LLMSs against these attacks.
We further conducted a detailed empirical analysis at different phases to uncover the preferences.
Finally, we propose seven \mseo strategies and conduct an end-to-end experiment based on the 450 self-deployment websites.
All identified issues were responsibly disclosed to major \LLMS vendors.

\noindent\textbf{Contribution.}
This work makes the following three contributions.

\begin{itemize}
    \item We provide a detailed investigation of real-world \LLMS products, uncovering their multi-phase workflows and identifying phase-specific attack surfaces.
    \item We reveal that the \LLMSs can resist over 99.78\% traditional black-hat SEO attacks, with the \textit{Retrieval} phase serving as the primary filter, intercepting the vast majority of malicious queries.
    \item We report that \LLMSs are vulnerable to \mseo attacks, i.e., rewritten-query stuffing and segmented text, double the manipulation rate compared to the baseline. 
    
\end{itemize}


%% file: section/2-Background.tex
\section{Background}


\subsection{\LLMS \& Black-Hat SEO}

\noindent\textbf{\LLMS.}
%
%
%
%
\cp{
LLM-enhanced search engines~(\LLMSs), also known as AI-powered search, combine real-time retrieval with generative summarization and are now widely adopted. Perplexity reports 169M monthly visits~\cite{similarweb}, and ChatGPT officially added search capabilities in 2024~\cite{krause2024}. Prior work has examined their efficiency, accuracy, and verifiability~\cite{spatharioti2023comparing, liu2023evaluating, wu2024easily, luo2025unsafe}, while adversarial studies explored visibility manipulation through GEO~\cite{aggarwal2024geo} and prompt injection~\cite{nestaas2024adversarial, pfrommer2024ranking}. 
}



\noindent\textbf{Black-Hat SEO.}
%
%
%
\cp{
Search Engine Optimization~(SEO) refers to improving website ranking and organic traffic through legitimate means such as optimizing structure, content, and user experience. In contrast, black-hat SEO manipulates rankings by violating search engine guidelines, aiming for short-term gains through techniques such as link farms~\cite{Wu2005linkfarm}, keyword stuffing~\cite{Ntoulas2006stuffing}, search redirection~\cite{Nektarios2011redirection, leontiadis2014nearly}, 
\ww{cloaking~\cite{wu2005cloaking, wang2011cloak}, semantic confusion} through ad injection or jargon obfuscation~\cite{liao2016seeking, yang2021scalable}, and long-tail keyword attacks~\cite{liu2021characterizing, liao2016characterizing}. 
%
}

\subsection{Threat Model}


Motivated to promote specific websites, the attacker deliberately modifies the structure or content of the websites \ww{so they are favorably indexed by search engines.}. When a victim user inputs certain queries, \LLMS may surface these websites and incorporate the link to the untrusted website into the generated responses.

\noindent\textbf{Attacker's Goal.}
The attacker’s objective is to induce \LLMSs to embed attacker-controlled URLs within their responses. Since link-bearing outputs directly guide users to promoted sites, our analysis focuses on responses containing clickable references to attacker-controlled domains (\eg citations).



\noindent\textbf{Attacker's Capability.} 
We assumed that the attacker controls multiple websites and has full authority to customize both content and structure. 
However, the attacker has no access to the intermediate outputs of the \LLMS.

%% file: section/3-Ecosystem.tex
\section{Attack Surface Analysis of \LLMS}\label{sec:workflow}

\begin{figure*}[ht!]
    \centering
    \includegraphics[width=0.95\linewidth]{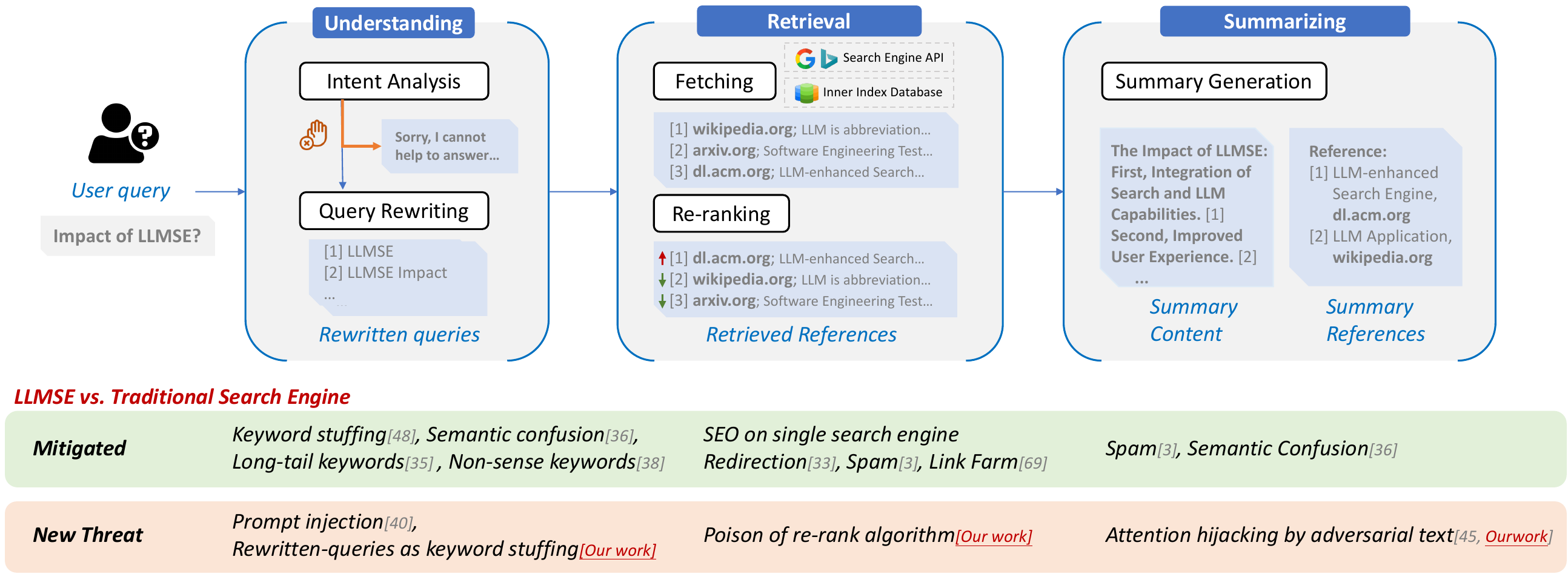}
    \caption{The Workflow and the Attack Surfaces of \LLMS. \cp{It includes three phases: (1) \textit{Understanding}: \LLMS analyzes user query's intent and rewrites the original query; (2) \textit{Retrieval}: fetching information from multiple databases with the rewritten queries, and re-ranking references; (3) \textit{Summarizing}: gathers all the structured information and generates the final answer, outputting the summary content and references.}}
    \Description{The Workflow and the Attack Surfaces of \LLMS.}
    \label{fig:llms-workflow}
\end{figure*}


In this section, we surveyed the current popular \LLMS systems across both open-source and closed-source markets. 
Through practical analysis, 
we can uncover the attack surface of each phase.


\subsection{Representative \LLMS Collection}

To gain a comprehensive understanding of the \LLMS ecosystem, we systematically collected a list of actively deployed \LLMSs from both industrial and open-source \ww{platforms}. 
%
\cp{
First, we searched keywords such as ``LLM search'' and ``AI search'' via Google. We extracted products from the top 100 search results and selected the five most frequently mentioned \LLMS products, which together account for over 98\% of the market~\cite{gs_statcounter}, representing those with the highest visibility and usage.
Second, we surveyed popular open-source repositories in ``LLM search'' and selected the top five \LLMS projects with over 10k stars from Github~\cite{felladrin2023awesome}, reflecting strong community adoption.
In total, we identified 10 representative \LLMSs, whose identities and popularity are summarized in Table~\ref{tab:llms_top}.
}



\input{table/llms-top}

\subsection{Attack Surface Analysis}

\cp{
\ww{
We conduct a comprehensive analysis on the collected \LLMS to uncover \LLMS attack surfaces.}  
For closed-source \LLMS, we systematically reviewed their homepage descriptions and official documentation, and manually interacted with them to observe user-facing outputs. 
For open-source \LLMS, we deployed them locally and analyzed their outputs and server-side logs to expose intermediate processing.
%
The workflow and the corresponding attack surface of \LLMS are summarized in Figure~\ref{fig:llms-workflow}.
}


\noindent\textbf{Phase 1: Understand the Query.} 
\LLMS understands the user's input 
, and provide actionable guidance for subsequent phases. This process contains two main components:


(1.1) \textit{Intent Analysis}: 
Infer user intent and decide whether external retrieval is needed~(\eg ChatGPT~\cite{chatgpt2024api}, Gemini~\cite{google2024grounding}). 
%
This step filters irrelevant or malicious parts of the query, helping defend against attacks like irrelevant keyword stuffing and semantic confusion. 
However, LLM-driven inference remains susceptible to adversarial attacks such as prompt injection or jailbreaks~\cite{liu2024formalizing}.

(1.2) \textit{Query Rewriting}: 
Rewrite the user input into one or more standardized queries~\cite{openui2024docs}. 
Among the examined \LLMSs, 8/10 clearly indicate that they actively regenerate queries,
often adopting a role-playing strategy~\cite{perplexica2024prompt,gptresearcher2023prompt}.
This step helps normalize phrasing and defend against adversarial manipulations based on typographic variations or misleading phrasing, \eg the long-tail SEO~\cite{liao2016characterizing} and non-sense keyword SEO~\cite{liu2025nokes}.
However, it might be exploited if attackers can predict rewritten queries and tailor stuffing attacks accordingly, as further discussed in Section~\ref{sec:attack}.

\noindent\textbf{Phase 2: Retrieve the Information.}
In this phase, the \LLMS executes the rewritten queries to gather candidate retrieved references and their content for subsequent processing.

(2.1) \textit{Fetching}: Employ external engines (\eg Google~\cite{google_custom_search_json_api}, Tavily~\cite{tavily2025api}) or inner database to fetch potentially relevant results.
Some \LLMSs can restrict the search scope to curated domains~\cite{perplexity2024guide,chatgpt2024search} to improve reliability.
The diversity of retrieval sources helps mitigate single-source poisoning.


(2.2) \textit{Re-ranking}: Scoring each retrieved content~\cite{google2024grounding} and filter for relevance.
The re-ranking process filters out spam-driven SEO abuse such as link farms, but its underlying scoring heuristics may unintentionally bias page selection, which we further examine in Section~\ref{sec:perf-retrieval} and explore its attack implications in Section~\ref{sec:attack}.

\noindent\textbf{Phase 3: Summarize the Answer.}
After retrieval, \LLMS synthesizes the summary and typically includes in-text citations or references to enhance credibility.


(3.1) \textit{Summary Generation}: Summarize a coherent and logically consistent response. Apart from ChatGPT and Gemini, most \LLMSs rely on external LLM APIs for content synthesis~\cite{openui2024docs}, with some integrating multiple LLMs~\cite{gptresearcher2023docs}.
It improves factual consistency and filters out spam or semantic-confusion content, yet remains vulnerable to adversarial text that can hijack model attention~\cite{nestaas2024adversarial}.

%% file: table/llms-top.tex
\begin{table}[t]
\centering
\footnotesize
\begin{threeparttable}
\caption{Overview of Representative \LLMSs}
\label{tab:llms_top}

    \begin{tabular}{llrcc}
    \toprule
    \textbf{\LLMS} & \textbf{Provider} &  \textbf{Popularity} & \textbf{Pub. Date} & \textbf{API} \\ 
    \midrule
    \multicolumn{5}{c}{Closed Source} \\ 
    \midrule
    ChatGPT Search~\cite{openai_chatgpt_search} & OpenAI &  5.91B Visits & 2024-10-31  & Y \\ 
    Gemini Grounding~\cite{geminiGoogle} & Google &  1.06B Visits & 2024-10-31 & Y \\ 
    Google AI Overview\cite{google_ai_overviews} & Google & 88.5B Visits & 2024-05-14 & N  \\ 
    Perplexity~\cite{perplexity} & Perplexity & 169M Visits& 2022-12-07 & Y \\ 
    Komo AI~\cite{komoai} & Komo & 201k Visits & 2023-01-18 & Y \\ 
    \midrule
    \multicolumn{5}{c}{Open Source} \\
    \midrule
    Open WebUI~\cite{openwebui2023} & Open WebUI &  112k Stars& 2023-10-07 & Y \\ 
    Khoj~\cite{khojai2023} & Khoj AI & 31.3k Stars& 2021-04-04 & N \\ 
    Storm~\cite{storm2024} & Stanford OVAL & 27.5k Stars & 2024-04-09 & Y\\ 
    Perplexica~\cite{perplexica2023} & ItzCrazyKns & 26.4k Stars& 2024-04-09 & Y \\ 
    GPT Researcher~\cite{gptresearcher2023} & Assaf Elovic &23.7k Stars& 2023-05-12 & Y \\ 
    \bottomrule
    \end{tabular}

\begin{tablenotes}
\footnotesize
\item 1) Monthly visits counts are from SimilarWeb~\cite{similarweb}; \# of stars are from Github.
\item 2) Google AI Overview is an internal experiments module of Google Search.
\item 3) All statistics are collected as of September 2025.
\end{tablenotes}
\end{threeparttable}


\end{table}

%% file: section/4-Traditional-Eva.tex
\section{Resilience of Black-Hat SEO Attack}
\label{sec:traditional-eva}



\cp{
Since \LLMSs are similar to traditional search engines, attackers intuitively apply existing black-hat SEO techniques to them. To assess how traditional black-hat SEO affects \LLMSs and dissect how the multi-phase mechanisms mitigate or amplify these manipulations,
%
we conduct a comprehensive evaluation of \LLMS resilience with a large-scale real-world black-hat SEO attacks in this section.
}

\subsection{Experiment Setup }\label{sec:seo-collect}

\cp{
To evaluate the resilience of \LLMSs against black-hat SEO attacks, we use Google Search as a representative traditional search engine to collect real-world successful attacks. 
An attack succeeding on Google but failing on an \LLMS indicates resilience.
All samples from existing attacks ensure both ethical compliance and diversity. 
}





\noindent\textbf{SEO-Bench Construction.} 
\cp{
To find black-hat SEO websites, we first conducted a literature review, getting five categories of black-hat SEO attack techniques that have well-established definitions: 
\solidnum{1} \textit{Semantic Confusion}: blends copied legitimate text with illicit promotions to dilute malicious intent, which raises ranking and evades filters; \solidnum{2} \textit{Redirection}: exploits vulnerabilities on high-authority sites to forward users to promoted targets, thereby inheriting the trusted site's credibility; \solidnum{3} \textit{Cloaking}: detects crawlers via request headers and serves SEO-optimized content to search engines while presenting different promotional or unrelated content to real users; \solidnum{4} \textit{Keyword Stuffing}: embeds excessive or trending terms to inflate apparent relevance and manipulate ranking algorithms; \solidnum{5} \textit{Link Farm}: creates large interlinked networks of low-quality sites to artificially raise link-based authority scores. 
}
Based on these studies, we reproduced the classification methods proposed in the corresponding works and tuned the respective classifiers. 
The implementation and evaluation are provided in Appendix~\ref{sec:classifier}.

Then, we selected appropriate origin keyword queries, including illegal-words and hot-words.
Illegal-words typically involve terms related to illegal or prohibited content, reflecting the underlying incentives for black-hat SEO; we extracted 2,499 illegal-words from prior studies~\cite{wang2011cloak, leontiadis2011measuring, liao2016seeking, yang2021scalable, zhang2024into}. 
Hot-words include popular search terms unrelated to the actual page content, which attackers use to boost visibility in rankings; we collected 9,301 hot-words from Google Trends~\cite{google_trends} over a six-month period (Nov. 2024 – Apr. 2025).
We then queried these 11,800 keywords on Google, and saved the top 50 search results, including their titles, summaries, URLs, redirection chains, and HTML content.
From over 500M collected websites, we identified 1,602 valid \texttt{query-website} pairs by the classifiers.
%
To ensure a balanced representation of each attack type in the dataset, we selected 200 pairs for each of the five SEO attack categories. As a result, our SEO-Bench dataset consists of a total of 1,000 \texttt{query-website} pairs. 
Table~\ref{tab:seo-attack-intro} shows the dataset details.

\input{table/seo-attack-intro}


\input{table/llms-seo-new}
\noindent\textbf{\LLMS Defense Evaluation.}
\cp{
We evaluate nine \LLMSs introduced in Section~\ref{sec:workflow}, excluding Google AI Overview due to its limited and unstable availability~\cite{googleaioverview2024}.
%
For closed-source \LLMSs, we select their first version with full search functionalities, i.e., \texttt{gpt-4o-mini}, \texttt{gemini-1.5}, \texttt{sonar}, \texttt{komo}.
For open-source \LLMSs, we deploy them locally with default configurations, and use \texttt{gpt-4o-mini} as the summarization model to ensure comparability across systems.
}

To quantify the resilience of \LLMSs against black-hat SEO attacks, we evaluate their ability to block target websites across different phases. Each entry in SEO-Bench is a query–website pair $(q_i, t_i)$, where $q_i$ is a search query and $t_i$ is the associated black-hat SEO website. 
We independently evaluate the resilience of each phase using a phase-specific blocking rate, defined as the proportion of attacks intercepted at that phase among those entering the phase. 
Specifically:
%
%
In \textit{Understanding} phase, an attack is blocked if the \LLMS decides not to proceed with retrieval after analyzing, indicating an early rejection of the search. 
In \textit{Retrieval} phase, an attack is blocked if the \LLMS performs retrieval but the SEO website does not appear in the retrieved results.
In \textit{Summarizing} phase, an attack is blocked if the SEO website appears in the retrieval reference but is excluded from the final reference. 
We also employ \textit{Cumulative Resilience} to intuitively capture the overall interception achieved after each phase.
The metrics are in Appendix~\ref{sec:metrics}.

\subsection{Landscape}\label{sec:finding}

We assess nine \LLMSs with three trials per query (27,000 requests).
Results are summarized in Table~\ref{tab:llms-seo-new}.

Our evaluation shows that \LLMSs are highly effective against black-hat SEO attacks, with the \textit{Understanding}, \textit{Retrieval}, and \textit{Summarizing} phases blocking 15.7\%, 98.2\%, and 85.2\% of attacks at their respective phases. Overall, they achieve a cumulative blocking rate of 99.78\%, where \textit{Retrieval} plays the most decisive role by filtering the majority, and \textit{Summarizing} adds a strong safeguard before output generation.
These results highlight the importance of layered defenses in LLM-based systems, enabling them to significantly outperform traditional search engines in resisting SEO attacks.


Although the results show that black-hat SEO techniques can still influence \LLMSs, 
the resilience varies significantly across \LLMSs.
ChatGPT is not affected by any black-hat SEO attack with a high refusal rate. Notably, open-source \LLMSs keep great defense performance, due to the fact that they choose the search engine API such as Tavily~\cite{tavily2025api} or SearXNG~\cite{searxng2025api}, \ww{which} \cp{provide} optimized source. In contrast, Komo and Perplexity are most affected.

The impact of different types of black-hat SEO attacks on \LLMSs varies as well. Semantic Confusion and cloaking pose the greatest risks in the final output to \LLMSs. For example, Komo is severely affected by Semantic Confusion, with a low filtering of 83.0\% and 73.5\%. Meanwhile, Gemini is only affected by Semantic Confusion. In contrast, although both redirection and cloaking attacks have successfully passed the \textit{Retrieval} phase on some \LLMSs, few of them advanced to the \textit{Summarizing} phase, thus failing to achieve a successful attack on the \textit{Summarizing} phase. Besides, the Keyword Stuffing poses no influence on any \LLMS.

\finding{\LLMSs exhibit strong resilience to black-hat SEO attacks, achieving the cumulative blocking rate of 99.78\%. Retrieval phase intercepts the vast majority of malicious queries. Semantic Confusion poses the greatest risks to \LLMSs.}

\subsection{Resilience on Understanding Phase}\label{sec:perf-understand}

Firstly, we measure how \LLMSs interpret and rewrite the query in the \textit{Understanding} phase helps mitigate black-hat SEO attacks.

\noindent\textbf{Intent Interception.}
%
\cp{
Upon receiving a user query, \LLMSs infer its intent to decide whether a web search is necessary. 
We analyze the interception mechanism by inspecting the specific API field values, as ChatGPT and Gemini explicitly indicate whether ``web\_search'' is invoked. 
Then we manually check these refused queries and corresponding answers.
In our result, 75.8\% of queries on ChatGPT are intercepted with no reference. Among them, 30.6\% are refused due to violations of safety policies, and 69.4\% are skipped due to intent misinterpretation, where the system treats the input as a statement rather than a search query.
This is because, unlike dedicated search engines, \LLMSs such as ChatGPT, which treat search as an auxiliary function, tend not to invoke search when they can answer based on internal knowledge. 
Similarly,  Gemini intercepts 16.3\% of queries, with 67.4\% refusals and 32.6\% misinterpretations. 
%
Although intent interception is not designed to counter SEO attacks, it can incidentally filter harmful or illicit queries before search execution, thereby reducing exposure to malicious content.
}



\finding[find:intent]{Intent interception enables \LLMSs such as ChatGPT to filter out 75.8\% of queries, effectively disrupting malicious SEO attempts at the start, even though unintentionally.}

\begin{figure}[t]
    \centering
    \includegraphics[width=0.9\linewidth]{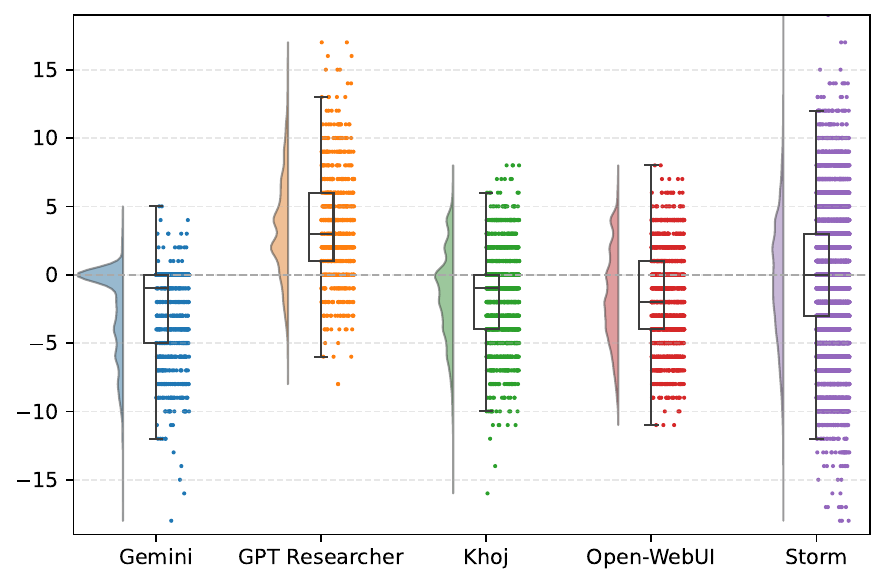}
    \caption{Distribution of Word Counts after Rewriting.}
    \Description{Distribution of Word Counts after Rewriting.}
    \label{fig:rewrite_length}
\end{figure}

\noindent\textbf{Query Rewriting.}
Before performing a real search in the \textit{Retrieval} phase, some \LLMSs generate a refined version of the original query. 
%
To investigate how this step influences the effectiveness of SEO-based manipulations, we extract the rewritten keywords on five of the collected \LLMSs 
that provide the rewritten queries in their API response, and compare them with the original inputs. 
Our analysis reveals that almost all of them prefer to rewrite the query. Figure~\ref{fig:rewrite_length} shows the changes in word count during the rewriting. 
For example, GPT Researcher shows a strong tendency to expand the query~(77.56\%), while others are more likely to shorten it.
Further examination of the system prompts indicates that the rewriting mechanisms, such as prefix/suffix modifications, query formatting, and targeted semantic enrichment, are guided by instructions aimed at improving search accuracy and user experience.


\cp{
To further examine how rewriting can mitigate the influence of black-hat SEO, we conducted a supplementary validation experiment on query rewriting.
As detailed in Appendix~\ref{sec:rewrite-eval}, reissuing rewritten queries to Google Search showed that 98.16\% failed to retrieve the original SEO websites, with even minor edits~(edit distance below 0.1) reducing attack success rates to under 10\%, confirming the strong disruptive impact of rewriting on adversarial rankings.
These results indicate that even slight syntactic modifications can substantially suppress exposure, suggesting that query rewriting serves as an effective and lightweight countermeasure.
}

\finding[find:rewrite]{
Query rewriting by \LLMSs effectively disrupts SEO attacks, including Long-tail and Keyword Stuffing. Even small edits can reduce the attack success rate to under 10\%.}

\subsection{Resilience on Retrieval Phase}\label{sec:perf-retrieval}

Then, we measure how fetching and re-ranking in the \textit{Retrieval}  phase helps mitigate black-hat SEO attacks.

\begin{figure}[t]
    \centering
    \includegraphics[width=0.95\linewidth]{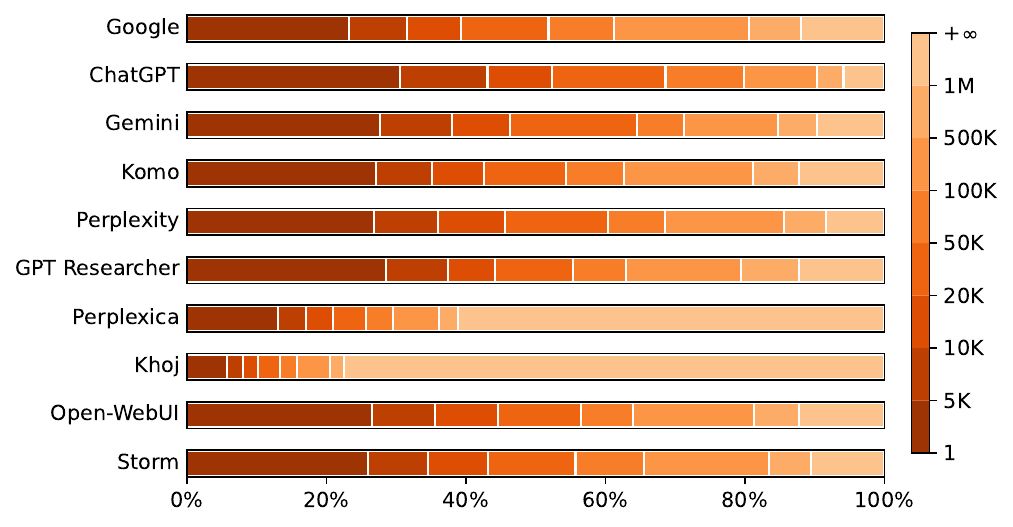}
    \caption{Domain Rank Proportions of Retrieved References from Both Traditional Search Engines and \LLMSs.}
    \Description{Domain Rank Proportions of Retrieved References from Both Traditional Search Engines and \LLMSs.}
    \label{fig:rank}
\end{figure}

\noindent\textbf{Fetching Preferences.}
\cp{
\LLMSs can restrict the search scope during fetching, so we examine whether their returned links are of higher quality than those from traditional engines. Using domain rankings~\cite{LePochat2019tranco}, we analyze retrieved references collected from trending queries. 
As shown in Figure~\ref{fig:rank}, most \LLMSs favor higher-ranked domains, with top-5k links appearing more frequently than lower-ranked ones. For instance, ChatGPT's share of authoritative links is nearly 50\% higher than Google’s. In contrast, \LLMSs such as Khoj and Perplexica exhibit over 60\% unreliable links due to hallucinated URLs. \ww{However}, while prioritizing authoritative sources improves result reliability and reduces SEO risks, excessive trust in high-ranking domains may introduce new vulnerabilities, such as malicious redirects or comment-based attacks on reputable sites.
}

\finding[find:retrieval]{\LLMS fetching preferences for authoritative websites enhance the overall search quality, but also emphasize the risk of compromised high-ranking sources.}

\noindent\textbf{Re-ranking Preferences.}
After fetching the content of web pages, \LLMSs often re-rank them based on internally defined quality assessment criteria. 
%
To investigate this preference, we examine the rank shifts of links that appear in both Google search results and \LLMSs. For each link that occurs in both result sets, we compute its relative rank change within the respective systems as 
\begin{equation}
\mathit{\Delta RelRank}_i = \mathit{Rank}_{\mathit{LLM}}^i - \mathit{Rank}_{\mathit{Google}}^i
\end{equation}
where $\mathit{Rank}_{\mathit{Google}}^i$ denotes the relative rank of the $i$-th link within the set of overlapping links in the Google search results, and $\mathit{Rank}_{\mathit{LLM}}^i$ denotes its relative rank in the \LLMSs results. To further uncover the re-ranking criteria employed by \LLMSs, we categorize websites with increasing relative rankings (i.e., $\Delta \mathit{RelRank}_i < 0$) as \textit{up} sites, and those with decreasing relative ranks (i.e., $\Delta \mathit{RelRank}_i > 0$) as \textit{down} sites. Then, we analyze several common features of the websites by computing the average values and their corresponding rates of differences, as summarized in Table~\ref{tab:rerank-feature}.




The results indicate that websites with increased rankings in \LLMSs typically exhibit a higher degree of text fragmentation (+19.04\%) and a greater presence of multimodal resources (+18.71\%) in terms of content. Structurally, they tend to feature denser internal linking (+14.89\%) and greater DOM depth (+11.36\%), which reflects a higher level of formatting complexity. These observed differences suggest a set of potentially influential factors that may reflect the preferences of \LLMSs. We will further examine their actual impact through controlled experiments in Section~\ref{sec:attack}.

\finding[find:rerank]{During the re-ranking, \LLMSs tend to favor webpages with higher content quality and content richness.}

\input{table/re-rank}
\input{table/illegal-propotion}

\subsection{Resilience on Summarizing Phase}\label{sec:perf-summary}

Finally, we measure summary intercepting of \textit{Summarizing} phase.

\noindent\textbf{Summary Interception.}
\cp{
To evaluate \LLMS resilience against illicit promotion during summary generation, we examine their ability to filter malicious content from illegal queries. We use an illicit-website classifier (Appendix~\ref{sec:classifier}) to measure the proportion of illegal references in \textit{Retrieval} and \textit{Summary} phases. Additionally, we analyze the semantics of the responses to determine how many illicit links contaminate the generated summaries.
As shown in Table~\ref{tab:illegal-proportion}, \LLMSs increasingly block malicious content as generation proceeds. In the \textit{Summary} phase, they remove on average 75.3\% more illegal links than in the \textit{Retrieval} phase, with Open-WebUI achieving the largest reduction~(97.9\%). 
Furthermore, 70.8\% illicit content is intercepted from the summary reference to the answer content.
These results indicate that \LLMSs generally favor neutral or positive content and actively suppress outputs involving violence, pornography, or other harmful material, reflecting their built-in safety mechanisms and alignment with normative standards.
}

We further analyze intercepted SEO attacks in the \textit{Summarizing} phase, including \textit{Semantic Confusion} and \textit{Cloaking}. Such pages often contain irrelevant or mixed content, weakening semantic relevance to the query. For example, confusion attacks may embed illicit promotions within otherwise legitimate text to enhance credibility. Appearing benign, they diverge from the illegal intent and are excluded from summaries. This suggests that, beyond rejecting harmful content, \LLMSs prioritize sources semantically aligned with query intent, which further mitigates SEO attacks.

\finding[find:summary]{\LLMSs prefer benign and semantically aligned content in summarization, refusing additional 75.3\% illegal websites. This strategy further mitigates the threat of illicit content and the impact of attacks such as Semantic Confusion.}

%% file: table/seo-attack-intro.tex
\begin{table}[t]
\centering
\footnotesize
\caption{Details of Black-Hat SEO Attacks in SEO-Bench}
\resizebox{\linewidth}{!}{
    \begin{tabular}{llll}
    \toprule
    \textbf{Black-Hat SEO Attack} & \textbf{Query} & \textbf{Classification Method} & \textbf{\#}\\ 
    \midrule
    Semantic Confusion~\cite{liao2016seeking,yang2021mingling,yang2021scalable} & Illegal-words & SCDS~\cite{yang2021mingling} & 200 \\
    Redirection~\cite{wang2007spam,leontiadis2011measuring,leontiadis2014nearly,mek2014detecthttp} & Illegal-words & Rule-Based Detector~\cite{leontiadis2011measuring} & 200 \\
    Cloaking~\cite{wu2005cloaking,niu2007quantitative,wang2011cloak,in2016cloakvisibility} & Illegal-words & Dagger~\cite{wang2011cloak} & 200 \\
    Keywords Stuffing~\cite{ntoulas2006detecting,araujo2010web,lu2011surf,zhang2014dspin} & Hot-words & Rule-Based Detector~\cite{ntoulas2006detecting} & 200 \\
    Link Farm~(SSP)~\cite{Wu2005linkfarm,chung2009study,john2011deseo,du2016seoever} & Hot-words & DNS Scanner~\cite{du2016seoever} & 200 \\
    \bottomrule
    \end{tabular}
}
\label{tab:seo-attack-intro}
\end{table}

%% file: table/llms-seo-new.tex
\begin{table*}[ht]
\centering
\footnotesize
\caption{Performance of Black-Hat SEO Attacks on \LLMSs. Each item: Resilience~(Und) / Resilience~(Ret) / Resilience~(Sum). 
}
\resizebox{\textwidth}{!}{
    \begin{tabular}{l|c|ccccc}
    \toprule
\textbf{\LLMS}
        & \textbf{Total Performance}
        & \textbf{Semantic Confusion}
        & \textbf{Redirection}
        & \textbf{Cloaking}
        & \textbf{Keywords Stuffing}
        & \textbf{Link Farm}\\

\midrule
ChatGPT  & 75.8\% / - / 100\%  & 92.5\% / - / 100\%  & 79.0\% / - / 100\%  & 84.0\% / - / 100\%  & 62.0\% / - / 100\%  & 61.5\% / - / 100\% \\
Gemini  & 16.3\% / - / 99.8\%  & 38.0\% / - / 99.0\%  & 4.5\% / - / 100\%  & 24.0\% / - / 100\%  & 3.0\% / - / 100\%  & 12.0\% / - / 100\% \\
Perplexity  & 4.6\% / 98.2\% / 64.7\%  & 12.0\% / 94.9\% / 55.6\%  & 1.5\% / 99.0\% / 50.0\%  & 6.0\% / 97.9\% / 100\%  & 0 / 100\% / -  & 3.5\% / 99.0\% / 50.0\% \\
Komo  & 4.2\% / 94.9\% / 67.3\%  & 0 / 83.0\% / 73.5\%  & 1.0\% / 99.0\% / 100\%  & 20.0\% / 93.8\% / 50.0\%  & 0 / 100\% / -  & 0 / 98.5\% / 33.3\% \\

    
Open-WebUI  & 8.1\% / 96.0\% / 100\%  & 12.1\% / 93.1\% / 100\%  & 0 / 100\% / -  & 7.7\% / 91.7\% / 100\%  & 0 / 100\% / -  & 17.9\% / 95.7\% / 100\% \\
Khoj  & 30.9\% / 99.6\% / 100\%  & 61.0\% / 100\% / -  & 19.5\% / 100\% / -  & 36.5\% / 100\% / -  & 16.5\% / 100\% / -  & 21.0\% / 98.1\% / 100\% \\
Storm  & 0 / 99.8\% / 50.0\%  & 0 / 100\% / -  & 0 / 100\% / -  & 0 / 100\% / -  & 0 / 100\% / -  & 0 / 99.0\% / 50.0\% \\
Perplexica  & 0 / 100\% / -  & 0 / 100\% / -  & 0 / 100\% / -  & 0 / 100\% / -  & 0 / 100\% / -  & 0 / 100\% / - \\
GPT Researcher  & 1.4\% / 95.5\% / 100\%  & 0 / 88.5\% / 100\%  & 0 / 97.5\% / 100\%  & 0 / 95.0\% / 100\%  & 3.0\% / 100\% / -  & 3.9\% / 98.0\% / 100\% \\
    \midrule
\textbf{Average Res.} & 15.7\% / 98.2\% / 85.2\% & 24.0\% / 95.4\% / 88.0\% & 11.7\% / 99.5\% / 90.0\% & 19.8\% / 97.6\% / 91.7\% & 9.4\% / 100.0\% / 100.0\% & 13.3\% / 98.7\% / 79.2\% \\
    \midrule
\textbf{Cumulative Res.} & 15.7\% / 98.48\% / 99.78\% & 24.0\% / 96.50\% / 99.58\% & 11.7\% / 99.56\% / 99.96\% & 19.8\% / 98.08\% / 99.84\% & 9.4\% / 100.00\% / 100.00\% & 13.3\% / 99.00\% / 99.96\% \\
    \bottomrule
    \end{tabular}
}
\label{tab:llms-seo-new}
\end{table*}

%% file: table/re-rank.tex
\begin{table}[t]
\centering
\footnotesize
\begin{threeparttable}
\caption{Different Features of Re-Ranked Websites}
    \begin{tabular}{lcccc}
    \toprule
    \textbf{Features} & \textbf{Avg.~(\textit{Up})} & \textbf{Avg.~(\textit{Down})} & \textbf{Differences} & \textbf{P-value}\\
    \midrule
    Text Fragmentation & 60.09 & 50.48 & \textbf{+19.04\%}  & \textbf{0.0100} \\
    DOM Depth & 13.93 & 12.51 & \textbf{+11.36\%} & \textbf{0.0036} \\
    Tag Diversity & 22.61 & 22.27 & +1.543\% & 0.8674 \\
    External Link \# & 14.71 & 15.81 & -6.971\% & 0.9901 \\
    Internal Link \# & 45.66 & 39.74 & \textbf{+14.89\%} & \textbf{0.0003} \\
    Multi-modal \# & 12.50 & 10.53 & \textbf{+18.71\%} & \textbf{0.0292} \\
    Meta Completeness & 0.4167 & 0.4196 & -0.6911\% & 0.8809 \\
    Alt Coverage & 0.2899 & 0.2873 & +0.9050\% & 0.8540 \\
    \bottomrule
    \end{tabular}
\label{tab:rerank-feature}
\begin{tablenotes}
\footnotesize
\item If p-value $< 0.05$, the difference is considered statistically significant~\cite{bruce2017practical}.
\end{tablenotes}
\end{threeparttable}
\end{table}

%% file: table/illegal-propotion.tex
\begin{table}[t]
\centering
\footnotesize
\caption{Illegal Proportion in Different Phases.}
\label{tab:illegal-proportion}
\begin{tabular}{lccc}
\toprule
\textbf{LLMSE} & \makecell{\textbf{Retrieved}\\ \textbf{References}} & \makecell{\textbf{Summary} \\ \textbf{References~($\delta$)}} & \makecell{\textbf{Summary} \\ \textbf{Content~($\delta$)}} \\
\midrule
ChatGPT        & --     & 2.00\% (--)          & 2.00\% (0.0\%) \\
Gemini         & --     & 0.00\% (--)          & 0.00\% (0.0\%) \\
Komo           & 4.36\% & 1.53\% ($\downarrow$ 64.9\%)     & 0.28\% ($\downarrow$ 81.7\%) \\
Perplexity     & 0.69\% & 0.35\% ($\downarrow$ 49.3\%)     & 0.00\% ($\downarrow$ 100.0\%) \\
GPT Researcher & 10.07\% & 3.48\% ($\downarrow$ 65.4\%)    & 1.55\% ($\downarrow$ 55.5\%) \\
Perplexica     & 1.24\% & 0.19\% ($\downarrow$ 84.7\%)     & 0.00\% ($\downarrow$ 100.0\%) \\
Khoj           & 1.27\% & 0.05\% ($\downarrow$ 96.1\%)     & 0.00\% ($\downarrow$ 100.0\%) \\
Open-WebUI     & 10.98\% & 0.23\% ($\downarrow$ 97.9\%)    & 0.00\% ($\downarrow$ 100.0\%) \\
Storm          & 1.61\% & 0.50\% ($\downarrow$ 69.0\%)     & 0.00\% ($\downarrow$ 100.0\%) \\
\midrule
\textbf{Avg.} & 4.89\% & 0.92\% ($\downarrow$ 75.3\%) & 0.43\% ($\downarrow$ 70.8\%) \\
\bottomrule
\end{tabular}
\end{table}

%% file: section/5-New-Attack.tex
\section{\mseo Attack}\label{sec:attack}

\input{table/attack-effectiveness}

\cp{
Building on these findings, we propose and evaluate novel end-to-end attacks tailored to \LLMSs, extending beyond prior work that focused only on ranking or summarization components~\cite{zou2024poisonedrag,aggarwal2024geo,Alaofi2024LLMsFooledRelevant}.
}

\subsection{\mseo Attack Techniques}

\cp{
As each phase of the \LLMS workflow~(\textit{Understanding}, \textit{Retrieval}, and \textit{Summarizing}) exposes distinct security risks, we design phase-specific \mseo attack strategies to systematically evaluate and manipulate these vulnerabilities.
}

\noindent\textbf{Attacking Understanding Phase.}
In this phase, 
rewriting introduces ambiguity and susceptibility to manipulation. To exploit this vulnerability, we design a targeted attack.

\begin{itemize}
    \item \textit{Rewritten-query Stuffing.} Embed potential rewritten queries extensively within web pages. Since most \LLMS may not always use the original search key for retrieval~(Section~\ref{sec:workflow}), predicting possible new queries and inserting them in web content can increase the likelihood of being matched.
\end{itemize}

\cp{
Note that while prompt injection is an important attack vector in this phase~(see Section~\ref{sec:workflow}), its impact is well studied~\cite{nestaas2024adversarial,pfrommer2024ranking,chen2025secalign,liu2024formalizing} and highly prompt-dependent, so we exclude it from our experiments.
}

\noindent\textbf{Attacking Retrieval Phase.}
Aiming to increase the priority in re-ranking and inspired by Finding~\ref{find:rerank}, we propose new techniques designed to align with the scoring mechanism.

\begin{itemize}
    
    \item \textit{Internal Links.} Embed numerous internal links within web pages to construct a network, simultaneously increasing the number of links to key pages.
    
    
    \item \textit{Multi-modal Resources.} Incorporate multi-modal resources~(\eg text, images, and videos) into web pages to increase their richness and boost perceived credibility. 

     \item \textit{Nested Structure.} Use structured labels and indexing that are better suited for retrieval to enhance the readability and accessibility of content,
     helping \LLMS quickly locate and extract key information, gaining an advantage in candidate selection. 

    \item \textit{Segmented Text.} Reduce the length of individual text segments. Shorter texts are often more suitable for direct citation.
    
\end{itemize}


\noindent\textbf{Attacking Summarizing Phase.}
To influence this phase, we design optimization strategies targeting both relevance and format.

\begin{itemize}
    \item \textit{Relevance Enhancement.} Focus on core keywords to enhance the semantic relevance and coherence of the text to the query. 
    
    \item \textit{Q\&A Formatting.} Present content in a question-and-answer~(Q\&A) format in the conversational tone of LLM-generated responses, increasing the likelihood of direct reuse by \LLMS.
\end{itemize}

\subsection{Effectiveness Evaluation}\label{sec:effective}

To examine the applicability of these \LLMS attack techniques influencing \LLMSs, we conducted a competitive experiment to compare the efficacy of various attack strategies under real-world conditions.

\noindent\textbf{Methodology.}
%
To evaluate the effectiveness of different \LLMS attack techniques, we deployed blog websites under a controlled domain, each promoting a different brand of the same type of product. We then queried \LLMSs for product recommendations using domain-restricted prompts (Appendix~\ref{sec:prompt}), and recorded the proportion of recommended sites associated with each attack technique.
\cp{This restriction reduces ethical concerns arising from real-world search pollution while enabling fair comparison across techniques.}
A higher occurrence suggests a stronger alignment between the corresponding manipulation and the preferences of the \LLMS. 


To mitigate the impact of randomness in \LLMS responses, the query was repeated 10 times per \LLMS, and we aggregated the total number of times each site appeared in the responses.
In addition to the seven \LLMS attack types, we included one non-SEO and one traditional SEO attack~(\ie Semantic confusion, which showed the best performance in Section~\ref{sec:finding}) for baseline comparison.
For each attack type, 50 websites were created. In total, the experiment involved 450 adversarial websites.
To ensure ethical compliance, all websites were labeled as ``For Testing Purposes Only'' and taken offline after the experiment finished. This ensured minimal long-term impact while maintaining the integrity of real-world testing.





\noindent\textbf{Implementation.}
We implemented these \LLMS attacks across various websites.
Specifically, we first generated a set of products using the same pattern, \ie ``Brand'' + ``Entity'' noun, and generated the base content with \texttt{gpt-4o-mini}. We also use the model for \textit{Rewritten-query Stuffing}, generating rewritten queries and embedding them into web pages.
For the \textit{Internal Links}, we embedded hyperlinks among the 50 websites of this type, forming mutual linking connections in the ``Useful Links'' block.
For the \textit{Multi-modal Resources}, we doubled the number of images in the base website content to increase visual richness.
In the \textit{Nested Structure}, we added an additional layer of subheadings, expanding from second-level to third-level headers to increase structural complexity.
For the \textit{Segmented Text}, we restructured the content by halving the average paragraph length, resulting in more segmented text blocks.
For the  \textit{Relevance Enhancement}, we removed the irrelevant part, \eg teams, and added more descriptions about products.
In the \textit{Q\&A Formatting}, each paragraph was prefaced with a question, followed by a corresponding answer block to simulate a question-answer format.
In the \textit{Semantic Confusion}, we inserted a promotional segment into unmodified news content.

\noindent\textbf{Results.}
Table~\ref{tab:attack-effective} presents the performance of \mseo attacks on \LLMSs, where each row shows the proportion of a specific attack type among all successful attacks.
\footnote{Notably, \texttt{gpt-4o-mini} and \texttt{gemini-1.5} refused to access all provided URLs in this experiment, likely due to stricter content-fetching policies~\cite{gemini-url-context}.
}
%
%
%
%
Overall, all proposed attacks demonstrated measurable effectiveness on \LLMSs, with each achieving performance above the baseline in at least one \LLMS.
%
In all attacks, attacks targeting the \textit{Retrieval} phase were more effective. 
\textit{Segmented Text} achieved the highest attack across most \LLMS platforms, with an exposure rate exceeding 50\% on Perplexica and GPT-Researcher, indicating that \LLMSs are better at understanding short and segmented content.
The second most effective technique was \textit{Rewritten-query Stuffing}, which doubled the exposure rate compared to the baseline in five \LLMSs, highlighting its strong influence on downstream outcomes.
%

\cp{
We further compare attack proportions across phases to assess each phase's filtering effects.
As \textit{Summarizing} proportions post-\textit{Retrieval} filtering, we focus on the attack effectiveness differences between the two phases to capture summary interception.
In Table~\ref{tab:attack-effective}, we mark significant increases in attack proportions during \textit{Summarizing}.
}
Content-driven strategies, such as \textit{Segmented Text} and \textit{Relevance Enhancement}, show noticeable growth, while \textit{Semantic Confusion}, as well as link- and resource-based tactics (\textit{Internal Links}, \textit{Multi-modal Resources}), tend to decline.
This shift suggests that, in \textit{Summarizing} phase, \LLMS is more influenced by content quality rather than the structure of external resources, underlining the importance of content-level manipulation.

%% file: table/attack-effectiveness.tex
\begin{table*}[ht]
\centering
\footnotesize
\caption{Success Rate of \mseo Attacks on \LLMSs. The percentage in each row indicates the success rate of a specific approach among all successful attacks. The bold numbers highlight the most effective attacks, and the up arrows ($\uparrow$) indicate the most improving attacks between \textit{Summarizing} phase and \textit{Retrieval} phase.}\label{tab:attack-effective}
\resizebox{\textwidth}{!}{
    \begin{tabular}{ll|cc|ccccccc}
    \toprule
    \multirow{3.5}{*}{\textbf{\LLMS}} 
        & \multirow{3.5}{*}{\textbf{Exposed Phase}} 
        & \multicolumn{2}{c|}{\textbf{Baseline}} 
        & \textbf{Understanding}
        & \multicolumn{4}{c}{\textbf{Retrieval}} 
        & \multicolumn{2}{c}{\textbf{Summarizing}} 
         \\
     \cmidrule(lr){3-4}  \cmidrule(lr){5-5}  \cmidrule(lr){6-9} \cmidrule(lr){10-11}  
        & & \textbf{Blank}
        & \makecell{\textbf{Semantic} \\ \textbf{Confusion} }
        & \makecell{\textbf{Rewritten-query} \\ \textbf{Stuffing} }
        & \makecell{\textbf{Internal} \\ \textbf{Links} }
        & \makecell{\textbf{Multi-modal}\\ \textbf{Resources} }
        & \makecell{\textbf{Nested} \\ \textbf{Structure}}
        & \makecell{\textbf{Segmented} \\ \textbf{Text} }
        & \makecell{\textbf{Relevance} \\ \textbf{Enhancement}}
        & \makecell{\textbf{Q\&A}\\ \textbf{Formatting} } \\
    \midrule
Perplexity & Retrieval & 7.29\% & 0.00\% & 19.79\% & 10.42\% & 8.33\% & 6.25\% & \textbf{28.12\%} & 11.46\% & 8.33\% \\ 
 & Summarizing & 0.00\% & 0.00\% & 25.00\% & 12.50\% & 12.50\% & 0.00\% & \textbf{37.50\%} $\uparrow$ & 12.50\% & 0.00\% \\ 
\midrule 
Komo & Retrieval & 8.33\% & 0.00\% & 23.96\% & 10.42\% & 9.38\% & 1.04\% & \textbf{30.21\%} & 10.42\% & 6.25\% \\ 
 & Summarizing & 2.44\% & 0.00\% & 26.83\% & 2.44\%  & 4.88\% & 0.00\% & \textbf{41.46\%} $\uparrow$ & 14.63\% & 7.32\% \\ 
\midrule 
Open-WebUI & Retrieval & 9.09\% & 2.60\% & 27.27\% & 2.60\% & 3.90\% & 11.69\% & \textbf{28.57\%} & 0.00\% & 14.29\% \\ 
 & Summarizing & 0.00\% & 0.00\% & 0.00\% & 0.00\% & 0.00\% & 0.00\% & 0.00\% & 0.00\% & 0.00\% \\ 
\midrule 
Khoj & Retrieval & 12.38\% & 3.47\% & 16.34\% & 4.95\% & 6.93\% & \textbf{26.73\%} & 17.33\% & 4.46\% & 7.43\% \\ 
 & Summarizing & 15.29\% & 1.18\% & 16.47\% & 4.71\% & 2.35\% & \textbf{30.59\%} & 22.35\% $\uparrow$ & 1.18\% & 5.88\% \\ 
\midrule 
Storm & Retrieval & 13.9\% & 1.36\% & \textbf{23.71\%} & 10.08\% & 9.26\% & 15.80\% & 9.54\% & 4.90\% & 11.44\% \\ 
 & Summarizing & 14.03\% & 0.45\% & 16.29\% & 10.86\% & 11.31\% & \textbf{19.91\%} $\uparrow$ & 8.14\% & 5.88\% & 13.12\% \\ 
\midrule 
Perplexica & Retrieval & 20.25\% & 0.00\% & 7.59\% & 5.06\% & 3.8\% & 7.59\% & \textbf{45.57\%} & 10.13\% & 0.00\% \\ 
 & Summarizing & 19.35\% & 0.00\% & 0.00\% & 6.45\% & 3.23\% & 6.45\% & \textbf{51.61\%} $\uparrow$ & 12.9\% & 0.00\% \\ 
\midrule 
GPT-Researcher & Retrieval & 0.00\% & 0.00\% & 21.74\% & 2.17\% & 6.52\% & 0.00\% & \textbf{65.22\%} & 0.00\% & 4.35\% \\ 
 & Summarizing & 0.00\% & 0.00\% & 24.32\% $\uparrow$ & 2.70\% & 8.11\% & 0.00\% & \textbf{59.46\%} & 0.00\% & 5.41\% \\ 
    \bottomrule
    \end{tabular}
}
\end{table*}

%% file: section/6-Disscussion.tex
\section{Discussion}

\noindent\textbf{Security Implications.}
The application of \LLMSs in information retrieval reshapes user trust and the threat landscape of black-hat SEO.
Users over-rely on \LLMS-generated summaries and references, perceiving them as authoritative, which magnifies the risks when malicious content and links are included in trusted outputs. 
Building upon traditional black-hat SEO, attackers are increasingly adapting their strategies to the internal preferences of \LLMSs, shifting content manipulation from isolated optimizations to system-level adversarial interactions. 
As \LLMS adoption grows, such weaknesses may gradually distort the Web's credibility structure, highlighting the need for timely, robust defenses to support a healthy information ecosystem and sustained user trust.

\noindent\textbf{Mitigation.}
To mitigate the vulnerabilities identified in this study, defenses for \LLMSs should be reinforced in a phase-aware manner across the entire workflow.
In the \textit{Understanding} phase, analyzing the stability of query rewriting and intent interpretation can help detect inputs deliberately aligned with rewriting behaviors; paraphrasing-based filtering, as shown in PoisonedRAG~\cite{zou2024poisonedrag}, can counter poisoning attempts in RAG systems.
During \textit{Retrieval}, mitigation should go beyond static domain authority by incorporating behavior-based signals, such as redirection patterns and cross-query reference consistency, to identify abused high-credibility sources.
In the \textit{Summarizing} phase, additional safeguards are needed against prompt- or text-based manipulations embedded in retrieved pages that may bias summarization preferences.
Furthermore, user awareness and transparency features, such as link provenance or credibility indicators, are crucial to reduce over-reliance on generated outputs and promote critical content verification, collectively enhancing the resilience of the \LLMS ecosystem.
\noindent\textbf{Limitation.}
\cp{
Our evaluation focused on ten representative \LLMSs selected by user scale and popularity, though other systems beyond this scope may demonstrate stronger resilience.
For ethical reasons, our implemented \mseo attacks were intentionally simplified and deployed for limited durations; real-world adversaries may employ more sophisticated or persistent methods, and the combined effects of multiple strategies remain unexplored.
}

%% file: section/8-Conclusion.tex
\section{Conclusion}

This work presents the first systematic security analysis of Large Language Model-enhanced Search Engines~(\LLMSs), revealing how black-hat SEO continues to influence their behaviors.
By analyzing phase-specific preferences and weaknesses, we demonstrate effective \mseo attacks that exploit these vulnerabilities.
We offer insights into more secure and resilient AI-driven search systems.

%% file: section/Acknowledge.tex
This work was supported by the New Generation Artificial Intelligence-National Science and Technology Major Project (No. 2025ZD0123204). 
Min Yang is a faculty of Shanghai Pudong Research Institute of Cryptology, Shanghai Institute of Intelligent Electronics \& Systems and Engineering Research Center of Cyber Security Auditing and Monitoring, and Shanghai Collaborative Innovation Center of Intelligent Visual Computing, Ministry of Education, China.

%% file: section/Appendix.tex
\appendix

\section{Black-Hat SEO Website Classifier}\label{sec:classifier}

\subsection{Implementation}
For five types of attacks, we replicated the methods from existing works~\cite{yang2021mingling,leontiadis2011measuring,wang2011cloak,ntoulas2006detecting,du2016seoever} for semi-automated detection. The specific implementation is as follows:

\noindent\textbf{Semantic Confusion.} 
We use two models to complete the task. (1) Context semantic classifier, used to predict the probabilities that a web page belongs to 14 benign topics, outputting \textit{prob\_14} (2) Malicious web page classifier, used to predict the probability that a web page is malicious, outputting \textit{prob\_{malicious}}.
Both models are based on the TextCNN architecture: Vocabulary Size = 10,000; Maximum Sequence Length = 500; Embedding Dimension = 128; Convolution Filter Sizes =[3,4,5]; Number of Filters per Size = 128; Pooling Layer = GlobalMaxPooling1D; Dropout Rate = 0.5; Optimizer = Adam; Batch Size = 64.

Judgment: 
\textit{max(prob\_14)} \textgreater 0.9 And \textit{prob\_{malicious}} \textgreater 0.9.

\noindent\textbf{Redirection.} 
We identify two types of redirection. (1) reputable domain redirecting to malicious content~( Illegal search keywords + Tranco Top 10,000 domain~\cite{LePochat2019tranco} or education/government domains + Redirection + Redirect to malicious website (2) benign search redirecting to malicious content~(Hot search keyword + Redirection + Redirect to malicious website). The malicious website classification model uses the same Malicious web page classifier in Semantic Confusion, outputting \textit{prob\_malicious}.

Judgment:
\textit{prob\_malicious} \textgreater 0.9.

\noindent\textbf{Cloacking.} 
We use the user agents of Google bot and users to crawl and obtain the page content of the two views. (1) Use text slicing techniques to generate content signatures and compare the similarity (\textit{signature\_sim}) between the user page and the bot crawled page; (2) Remove the blank pages; (3) Calculate the matching degree of the summary on the user page and the bot crawled page(\textit{summary\_sim}) (4) Calculate the DOM structure similarity (\textit{DOM\_sim}) of two views.

Judgment:
\textit{signature\_sim} \textless 0.9 And \textit{summary\_sim} \textgreater 0.33 And \textit{DOM\_sim} \textgreater 0.66.

\noindent\textbf{Keywords Stuffing.} 
We consider the keywords in Google Trends~\cite{google_trends}. (1) Compute the number of Google's hot search terms matched on the page, \textit{hotwords\_count}. (2) Use the “site:domain” query in Google to determine whether the number of sub-pages is very large and all are spam content. If both conditions are satisfied, then we consider the page \textit{Keywords Stuffing}.

Judgment:
 \textit{hotwords\_count}  \ensuremath{\ge} 10 And \textit{spam\_subpages}  \ensuremath{\ge} 100.

\noindent\textbf{Link Farm.} 
We conduct DNS queries supporting wildcards. Then visit the homepage or sitemap twice and extract the set of hyperlinks on it, and finally get URL set \textit{A} and URL set \textit{B}.

Judgment: $\max\left(\frac{\vert A - B\vert}{\vert A \vert}, \frac{\vert A - B\vert}{\vert B\vert}\right) \geq 0.2$


\subsection{Evaluation}
To assess the effectiveness of our classifiers used in SEO-Bench construction, we conducted a systematic evaluation for each of the five attack categories. This section outlines the evaluation methodology and presents the corresponding results.

For each classifier, we computed a confusion matrix based on manual verification. Specifically, we sampled 100 websites that were predicted as positive (\texttt{label=1}) and 100 websites predicted as negative (\texttt{label=0}). Each sample was manually checked to determine whether the prediction matched the black-hat SEO characteristics.
From these manual labels, we derived standard classification metrics including accuracy, precision, recall, and F1-score. 

The overall accuracy across all classifiers was 91.12\%, indicating sufficient reliability for use in dataset construction. 
Table~\ref{tab:confusion-matrics-1}-\ref{tab:confusion-matrics-5} presents the confusion matrices for the five classifiers, providing a detailed view of performance across different attack types.

\begin{table}[ht]
    \centering
    \caption{Evaluation Metrics for the Redirection classifier.}
    \label{tab:confusion-matrics-1}
    \footnotesize
    \begin{tabular}{l|cc|c}
        \toprule
        & \textbf{Predicted Positive} & \textbf{Predicted Negative} & \textbf{Total} \\
        \midrule
        \textbf{Actual Positive} & 99  & 21  & 120 \\
        \textbf{Actual Negative} & 1   & 79  & 80 \\
        \midrule
        \textbf{Total}      & 100      & 100      & 200 \\
        \midrule
        \multicolumn{2}{l}{
            \textbf{Accuracy:} 89.0\%
        } & \multicolumn{2}{l}{
            \textbf{Precision:} 99.0\%
        } \\
        \multicolumn{2}{l}{
            \textbf{Recall:} 82.5\%
        }& \multicolumn{2}{l}{
           \textbf{F1 Score:} 89.4\%
        } \\
        \bottomrule
    \end{tabular}
\end{table}

\begin{table}[ht]
    \centering
    \caption{Evaluation Metrics for the Cloaking classifier.}
    \label{tab:confusion-matrics-2}
    \footnotesize
    \begin{tabular}{l|cc|c}
        \toprule
        & \textbf{Predicted Positive} & \textbf{Predicted Negative} & \textbf{Total} \\
        \midrule
        \textbf{Actual Positive} & 87  & 5  & 92 \\
        \textbf{Actual Negative} & 13  & 95 & 108 \\
        \midrule
        \textbf{Total}      & 100      & 100      & 200 \\
        \midrule
        \multicolumn{2}{l}{
            \textbf{Accuracy:} 91.0\%
        } & \multicolumn{2}{l}{
            \textbf{Precision:} 87.0\%
        } \\
        \multicolumn{2}{l}{
            \textbf{Recall:} 94.6\%
        }& \multicolumn{2}{l}{
           \textbf{F1 Score:} 90.6\%
        } \\
        \bottomrule
    \end{tabular}
\end{table}

\begin{table}[ht]
    \centering
    \caption{Evaluation Metrics for the Keyword Stuffing classifier.}
    \label{tab:confusion-matrics-3}
    \footnotesize
    \begin{tabular}{l|cc|c}
        \toprule
        & \textbf{Predicted Positive} & \textbf{Predicted Negative} & \textbf{Total} \\
        \midrule
        \textbf{Actual Positive} & 89  & 0  & 89 \\
        \textbf{Actual Negative} & 11  & 100 & 111 \\
        \midrule
        \textbf{Total}      & 100      & 100      & 200 \\
        \midrule
        \multicolumn{2}{l}{
            \textbf{Accuracy:} 94.5\%
        } & \multicolumn{2}{l}{
            \textbf{Precision:} 89.0\%
        } \\
        \multicolumn{2}{l}{
            \textbf{Recall:} 100.0\%
        }& \multicolumn{2}{l}{
           \textbf{F1 Score:} 94.18\%
        } \\
        \bottomrule
    \end{tabular}
\end{table}

\begin{table}[ht]
    \centering
    \caption{Evaluation Metrics for the Semantic Confusion classifier.}
    \label{tab:confusion-matrics-4}
    \footnotesize
    \begin{tabular}{l|cc|c}
        \toprule
        & \textbf{Predicted Positive} & \textbf{Predicted Negative} & \textbf{Total} \\
        \midrule
        \textbf{Actual Positive} & 77  & 4  & 81 \\
        \textbf{Actual Negative} & 23  & 96  & 119 \\
        \midrule
        \textbf{Total}      & 100      & 100      & 200 \\
        \midrule
        \multicolumn{2}{l}{
            \textbf{Accuracy:} 86.6\%
        } & \multicolumn{2}{l}{
            \textbf{Precision:} 77.0\%
        } \\
        \multicolumn{2}{l}{
            \textbf{Recall:} 95\%
        }& \multicolumn{2}{l}{
           \textbf{F1 Score:} 87.74\%
        } \\
        \bottomrule
    \end{tabular}
\end{table}

\begin{table}[ht]
    \centering
    \caption{Evaluation Metrics for the Link Farm classifier.}
    \label{tab:confusion-matrics-5}
    \footnotesize
    \begin{tabular}{l|cc|c}
        \toprule
        & \textbf{Predicted Positive} & \textbf{Predicted Negative} & \textbf{Total} \\
        \midrule
        \textbf{Actual Positive} & 92  & 3  & 95 \\
        \textbf{Actual Negative} & 8   & 97 & 105 \\
        \midrule
        \textbf{Total}      & 100      & 100      & 200 \\
        \midrule
        \multicolumn{2}{l}{
            \textbf{Accuracy:} 94.5\%
        } & \multicolumn{2}{l}{
            \textbf{Precision:} 92.0\%
        } \\
        \multicolumn{2}{l}{
            \textbf{Recall:} 96.8\%
        }& \multicolumn{2}{l}{
           \textbf{F1 Score:} 94.3\%
        } \\
        \bottomrule
    \end{tabular}
\end{table}



\section{Evaluation Metrics}\label{sec:metrics}

\begin{equation}
    \mathit{Resilience(Und)}= \frac{\left|\{(q_i, t_i) \mid \text{\uquery}(q_i) = \emptyset \}\right|}{\left| \left\{ (q_i, t_i) \right\} \right| }
\end{equation}

\begin{equation}
\mathit{Resilience(Ret)} = \frac{ \left| \left\{ (q_i, t_i) \mid t_i \notin \text{\rlink}(q_i) \right\} \right| }{ \left|\{(q_i, t_i) \mid \text{\uquery}(q_i) \neq \emptyset \}\right| }
\end{equation}

\begin{equation}
\mathit{Resilience(Sum)} = \frac{ \left| \left\{ (q_i, t_i) \mid t_i \notin \text{\slink}(q_i) \right\} \right| }{ \left| \left\{ (q_i, t_i) \mid t_i \in \text{\rlink}(q_i) \right\} \right|  }
\end{equation}

\begin{equation}
Cumulative_k = \sum_{i=1}^{k} c_i, 
\quad \text{where}\
c_i = \left(1 - \sum_{j=1}^{i-1} c_j\right) \cdot Resilience_i.
\end{equation}

\section{Query Rewriting Validation Experiment}\label{sec:rewrite-eval}

To validate the effectiveness of query rewriting against SEO attacks, we resubmitted \uquery to Google Search and observed that 98.16\% failed to retrieve the original SEO websites, demonstrating the approach’s strong disruptive impact.
Then, we explore how the degree of rewriting affects retrieval by measuring semantic and syntactic differences using semantic textual distance~(STD) and edit distance~(ED).

Table~\ref{tab:rewrite-sts} shows that as semantic distance increases, the retrieval success rate drops significantly. Notably, even syntactic changes with an edit distance below 0.1 can reduce the success rate to under 10\%, and when semantic similarity remains high (STD $<$ 0.1), retrieval drops below 2\% if the edit distance exceeds 0.2.
Additionally, we observe 44 rewritten queries with a semantic distance greater than 0.5, indicating reversed or contradictory meanings~(cosine similarity $<$ 0).

This contrast suggests that query rewriting is effective not only by changing meaning but also by disrupting structural patterns used in SEO. Even minor edits can interfere with keyword matching while preserving the original intent, highlighting the role of rewriting in mitigating SEO attacks. 

\input{table/rewrite-sts}

\section{Domain-Restricted Query}\label{sec:prompt}

\begin{promptbox}{Prompt Example for Domain-Restricted Query}
Search strictly within site:\{domain\} for \{product\}, then recommend only results from this domain. Give me your answer and references.
\end{promptbox}

\section{Ethical Considerations}
This study is conducted under rigorous ethical oversight, and adheres to strict ethical guidelines to ensure responsible research practices.

(1) \textit{Controlled Experimentation.} All experiments were conducted in controlled environments to avoid real-world disruption.
As detailed in Section~\ref{sec:seo-collect}, we used existing SEO websites to build the dataset instead of generating new ones, preventing large-scale interference with real ecosystems~(\eg Google).
The \mseo attacks were implemented in simplified form and within a restricted scope, using controlled subdomains to avoid contaminating legitimate domains.
All test sites were clearly marked with ``For Testing Purposes Only'' disclaimers and taken down after experiments to eliminate residual impact.
For \LLMSs with limited search functions, such as ChatGPT and Gemini, we avoided further jailbreak attempts.

(2) \textit{Open Data.} No sensitive or private data was accessed in any experiment.
For closed-source \LLMSs, we strictly followed official API policies and interacted through default interfaces.
Open-source \LLMSs were deployed on isolated LAN servers using officially obtained API keys.
For SEO crawling, only publicly available Google Search results were requested, and Google-documented user agents~\cite{googleCommonCrawlers} were used to simulate crawler behavior.

(3) \textit{Responsible Disclosure.}
We adhered to the responsible disclosure. 
First, we reported to Google all 1,602 black-hat SEO websites identified in Section~\ref{sec:seo-collect} to facilitate timely remediation. 
Second, for the nine evaluated \LLMSs, we have contacted or are contacting both commercial and open-source providers through their official vulnerability disclosure channels (\eg, OpenAI's Bug Bounty) with detailed reports describing the vulnerabilities, reproduction steps, and suggested mitigation. 
For open-source systems, disclosures were or will be submitted via email to developers.
Third, for our temporary experimental blogs, removal requests were filed with Google and Bing to ensure de-indexing after experiment termination. 
These actions collectively demonstrate our commitment to responsibly exposing risks while assisting vendors in strengthening system resilience.

(4) \textit{Researcher Care.} We ensured the well-being of all researchers by providing methodological guidance and psychological support.
Given the potentially disturbing nature of illegal or harmful website content, annotators worked in a controlled and supportive environment with flexible schedules to prevent fatigue.
Regular mental health check-ins and access to counseling resources were maintained, and no participants reported psychological harm or undue stress during the study.

%% file: table/rewrite-sts.tex
\begin{table}[ht]
\centering
\footnotesize
\begin{threeparttable}
\caption{The Retrieval Success Rate at Different Semantic~(STD) and Syntactic~(ED) Changes}\label{tab:rewrite-sts}

    \begin{tabular}{cccccc}
    \toprule
    \textbf{ED$\backslash$STD} & \textbf{(0.0,0.1]} & \textbf{(0.1,0.2]} & \textbf{(0.2,0.5]} & \textbf{(0.5,1.0]} & \textbf{Total} \\
    \midrule
    (0.0,0.1] & \textbf{10.26\%} & 0.00\% & - & - & \textbf{9.52\%} \\
    (0.1,0.2] & 10.34\% & 0.00\% & 0.00\% & - & 9.55\% \\
    (0.2,0.5] & \textbf{1.49\%} & 1.91\% & 2.01\% & - & \textbf{1.64\%} \\
    (0.5,1.0] & 0.57\% & 0.71\% & 0.33\% & 0.00\% & 0.48\% \\
    \midrule
    Total & 2.46\% & 1.08\% & 0.57\% & 0.00\% & 1.46\% \\
    \bottomrule
    \end{tabular}
\begin{tablenotes}
\scriptsize
\item 1) STD: semantic textual distance, measured as $(1 - \text{cosine similarity}) / 2$;
\item 2) ED: edit distance, measured as $1 - \text{Levenshtein Ratio}$.
\end{tablenotes}
\end{threeparttable}
\end{table}